\documentclass[aps,preprint]{revtex4}%
\usepackage{amsfonts}
\usepackage{amsmath}
\usepackage{amssymb}
\usepackage{graphicx}%
\begin{document}
\preprint{ }
\title[ ]{Darwin-Lagrangian Analysis for the Interaction of a Point Charge and a Magnet:
Considerations Related to the Controversy Regarding the Aharonov-Bohm and
Aharonov-Casher Phase Shifts}
\author{Timothy H. Boyer}
\affiliation{Department of Physics, City College of the City University of New York, New
York, New York 10031}

\begin{abstract}
The classical electromagnetic interaction of a point charge and a magnet is
discussed by first calculating the interaction of a point charge with a simple
model magnetic moment and then suggesting a multiparticle limit. \ The Darwin
Lagrangian is used to analyze the electromagnetic behavior of the model
magnetic moment (composed of two oppositely charged particles of different
mass in an initially circular Coulomb orbit) interacting with a passing point
charge. \ Considerations of force, energy, momentum, and center of energy are
treated through second order in $1/c$. \ The changing magnetic moment is found
to put a force back on a passing charge; this force is of order $1/c^{2}$ and
depends upon the magnitude of the magnetic moment. \ The limit of a
many-particle magnet arranged as a toroid is discussed. \ It is suggested that
in the multiparticle limit, the electric fields of the passing charge are
screened out of the body of the magnet while the magnetic fields of the
passing charge penetrate into the body of the magnet. \ This is consistent
with our understanding of the penetration of electromagnetic velocity fields
into ohmic conductors. \ The proposed multiparticle limit is consistent with
the conservation laws for energy and momentum, as well as constant motion of
the center of energy, and Newton's third law for the net Lorentz forces on the
magnet and on the point charge. \ The work corresponds to a classical
electromagnetic analysis of the interaction which is basic to understanding
the controversy over the Aharonov-Bohm and Aharonov-Casher phase shifts and
represents a refutation of the suggestions of Aharonov, Pearle, and Vaidman.

\end{abstract}
\maketitle

\section{Introduction}

The interaction of a point charge and a magnet is a complicated and
controversial problem of electromagnetism. \ The problem is ignored by the
classical physics textbooks and is discussed in the research literature in
connection with the Shockley-James paradox,\cite{S-J} and in connection with
the Aharonov-Bohm\cite{AB} and Aharonov-Casher\cite{AC} phase shifts for
particles. \ The problem in understanding arises because the interaction
involves relativistic terms of order $1/c^{2}$ (where $c$ is the speed of
light in vacuum) which are not nearly so familiar as nonrelativistic
mechanics. Writing regarding the interaction of a point charge and a magnet in
1968, Coleman and Van Vleck remarked in an oft-cited article,\cite{CVV}
"Unfortunately, the equations which we have obtained are singularly resistant
to a simple physical interpretation in terms of particles exchanging forces;
..." \ However, despite the complications and in line with the controversy,
the problem is an important one which reflects back on our understanding of
classical electromagnetism and on the connections between classical and
quantum physics. \ 

\section{The Problem and the Controversy}

There are no electric or magnetic fields outside a long neutral solenoid or
toroid when the currents are maintained constant. \ Therefore \ when a charged
particle passes a long solenoid or a toroid, there are no electric or magnetic
fields at the position of the passing charge due to the \textit{unperturbed}
charge and current densities of the magnet. \ On the other hand, there are
clearly electric and magnetic fields due to the passing charge at the position
of the magnet. \ The electric fields of the passing charge will cause
accelerations of the charges which carry the currents which create the flux of
the magnet. \ Also, the magnetic fields of the passing charge will cause a net
Lorentz force on the magnet. \ Thus far the description would be approved by
all physicists. \ \ However, the response of the multiparticle magnet seems so
complicated that no one has calculated the magnet's response in detail. \ 

Since it does not seem possible at present to carry out a complete
multiparticle calculation starting from accepted theory, we are left with
suggestive partial calculations and hence with competing points of view
depending upon which aspects of the partial calculations are favored. \ At
present, there are two competing interpretations for the behavior of a magnet
and a passing point charge.

\subsubsection{The No-Velocity-Change Point of View}

The supporters of the quantum topological view\cite{AB}\cite{AC}%
\cite{APV}\cite{Vaidman}\cite{P} of the Aharonov-Bohm phase shift claim that
there are no velocity changes for the interacting charged particle or the
magnet. \ Indeed, the supporters of this view say that there are no
significant changes in the charge or current densities in the magnet.
\ Therefore the passing charge never experiences a Lorentz force and never
changes velocity. \ Furthermore, although the magnet does indeed experience a
net Lorentz force due to the magnetic field of the passing charge,
nevertheless the electric field of the passing charge penetrates into the
magnet giving \ a "hidden momentum in magnets" whose change "cancels" the net
magnetic Lorentz force on the magnet so that the center of energy of the
magnet is never disturbed. \ In this point of view, the electromagnetic fields
of the passing charge may cause confusion behind the scenes inside the magnet,
but there is no change in the magnet's center of energy and there is no
feedback signal sent to the passing charge which is causing the confusion in
the magnet.\cite{nature} \ 

\subsubsection{The Classical-Lag Point of View}

The classical-lag point of view\cite{Lieb}\cite{B2}\cite{B3}\cite{B4}%
\cite{B5}\cite{B6}\cite{B7}\cite{B8} takes a totally different perspective on
the changes in the charge and current densities induced in the magnet. \ In
this view, the induced densities lead to a Lorentz force back on the passing
charge which is equal in magnitude and opposite in direction to the net
magnetic Lorentz force which the magnetic field of the passing charge places
on the magnet. \ The electric charges on the surface of the magnet screen the
electric field of the passing charge out from the interior of the magnet, and
therefore there is no significant change in the momentum of the
electromagnetic fields. \ On the other hand, the magnetic field of the passing
charge penetrates into the magnet, and it is the magnetic energy change
associated with the overlapping magnetic fields which gives the magnitude of
the energy change of the passing charge due to the back force. \ This view
fits with what we know of the penetration of electric and magnetic velocity
fields into ohmic conductors. \ In this scenario, we have explicit ideas
concerning conservation of energy, linear momentum, and constant motion of the
center of energy. \ We also have the validity of Newton's third law for the
net Lorentz forces between the magnet and the passing charge.

Both points of view predict the Aharonov-Bohm and Aharonov-Casher phase
shifts. \ \ The no-velocity-change point of view claims that, in the light of
their interpretation, the phase shifts represent completely new quantum
topological effects occurring in the absence of classical forces, and there
are no classical analogues. \ The classical-lag point of view claims that the
phase shifts present classical velocity shifts analogous to those occurring
when only one beam of light passes through a piece of glass before two
coherent beams interfere. \ \ The conflict between the two points of view has
existed for thirty years without ever being put to experimental test to
determine whether or not there are velocity changes for the electrons passing
through a toroid or past a long solenoid. \ The no-velocity-change point of
view has been widely accepted because most physicists do not think of the
possibility of induced charge and current densities in magnets; they do
consider induced charge densities only in electrostatic situations.
\ Furthermore, the proponents of the no-velocity-change point of view have
declared that the lag point of view is impossible because i) the
electromagnetic fields of the passing charge would not penetrate into a
conductor surrounding a toroid or solenoid, and ii) the back electric field at
the passing charge could not be of order $1/c^{2}$ and proportional to the
magnetic flux of the magnet. \ The objection i) has been shown to be
groundless.\cite{B5} \ Magnetic velocity fields do indeed penetrate into good
conductors in exactly the required form which is completely different from the
exponential skin-depth form taken by electromagnetic wave fields.\cite{B1999} \ 

The objection ii) is addressed in the present article. \ In 1968 Coleman and
Van Vleck\cite{CVV} discussed the interaction of a stationary point charge and
a magnet using the Darwin Lagrangian. We will be following their approach in
the following analysis. \ We will discuss the interaction of a passing point
charge and a magnetic moment where the magnetic moment is modeled as a
classical hydrogen atom and where the electromagnetic interactions are carried
to order $1/c^{2}$ by using the Darwin Lagrangian. \ This is a well-defined
classical electromagnetic system which is relativistic through order $1/c^{2}%
$. \ In order to separate out the electrostatic effects (which are independent
of the magnetic moment) from magnetic effects dependent upon the magnetic
moment, we will sometimes average over atoms and anti-atoms with the same
magnetic moment. \ We will describe the motion and check all the conservation
laws. \ We will find that in this case the induced currents are important and
that there are electric Lorentz forces back on the passing charge which indeed
are of order $1/c^{2}$ and are proportional to the magnetic moment. \ There is
also a displacement of the center of energy of the magnetic moment. \ This
behavior contradicts the suggestions of the proponents of the
no-velocity-change point of view.\cite{APV}\cite{Vaidman} \ Next we will
discuss the passage to the limit of a multiparticle magnet. \ \ Finally, in
this multiparticle limit, we discuss the conservation-law aspects which are
mentioned above.

\section{The Darwin Lagrangian and Electromagnetic Fields}

The Darwin Lagrangian for particles of charge $e_{a}$, mass $m_{a}$,
displacement $\mathbf{r}_{a}$, and velocity $\mathbf{v}_{a}$ is given
by\cite{CVV}\cite{Jackson}%
\begin{align}
L  &  =%
{\displaystyle\sum\limits_{a}}
\left(  \frac{1}{2}m_{a}\mathbf{v}_{a}^{2}+\frac{1}{8c^{2}}m_{a}\mathbf{v}%
_{a}^{4}\right)  -\frac{1}{2}%
{\displaystyle\sum\limits_{a}}
{\displaystyle\sum\limits_{b\neq a}}
\frac{e_{a}e_{b}}{r_{ab}}\nonumber\\
&  +\frac{1}{2}%
{\displaystyle\sum\limits_{a}}
{\displaystyle\sum\limits_{b\neq a}}
\frac{e_{a}e_{b}}{2c^{2}r_{ab}}\left[  \mathbf{v}_{a}\cdot\mathbf{v}_{b}%
+\frac{(\mathbf{v}_{a}\cdot\mathbf{r}_{ab})(\mathbf{v}_{b}\cdot\mathbf{r}%
_{ab})}{r_{ab}^{2}}\right]
\end{align}
where $\mathbf{r}_{ab}=\mathbf{r}_{a}-\mathbf{r}_{b}$ and $r_{ab}%
=|\mathbf{r}_{a}-\mathbf{r}_{b}|$. \ Lagrange's equations of motion give a
canonical momentum%
\begin{equation}
\mathbf{p}_{a}^{canonical}=\frac{\partial L}{\partial\mathbf{v}_{a}}%
=m_{a}\mathbf{v}_{a}\left(  1+\frac{\mathbf{v}_{a}^{2}}{2c^{2}}\right)  +%
{\displaystyle\sum\limits_{b\neq a}}
\frac{e_{a}e_{b}}{2c^{2}r_{ab}}\left[  \mathbf{v}_{b}+\frac{\mathbf{r}%
_{ab}(\mathbf{r}_{ab}\cdot\mathbf{v}_{b})}{r_{ab}^{2}}\right]
\end{equation}
and a time derivative%
\begin{align}
\frac{d}{dt}\mathbf{p}_{a}^{canonical}  &  =\frac{\partial L}{\partial
\mathbf{r}_{a}}=%
{\displaystyle\sum\limits_{b\neq a}}
\frac{e_{a}e_{b}\mathbf{r}_{ab}}{2c^{2}r_{ab}^{3}}-%
{\displaystyle\sum\limits_{b\neq a}}
\frac{e_{a}e_{b}\mathbf{r}_{ab}}{2c^{2}r_{ab}^{3}}\left[  \mathbf{v}_{a}%
\cdot\mathbf{v}_{b}+\frac{3(\mathbf{v}_{a}\cdot\mathbf{r}_{ab})(\mathbf{v}%
_{b}\cdot\mathbf{r}_{ab})}{r_{ab}^{2}}\right] \nonumber\\
&  +%
{\displaystyle\sum\limits_{b\neq a}}
\frac{e_{a}e_{b}}{2c^{2}r_{ab}^{3}}\left[  \mathbf{v}_{a}(\mathbf{v}_{b}%
\cdot\mathbf{r}_{ab})+\mathbf{v}_{b}(\mathbf{v}_{a}\cdot\mathbf{r}%
_{ab})\right]
\end{align}
The Darwin Lagrangian accurately reflects the classical electromagnetic
interaction of charged particles through order $1/c^{2}$. \ To lowest order in
$1/c^{2},$ the interaction among the charges is given by the Coulomb force and
the nonrelativistic form of Newton's second law $\mathbf{F}=m\mathbf{a}$.
\ This $0$-order behavior can then be inserted back into the equations of
motion to allow calculation of the higher-order corrections. \ 

It is sometimes revealing to rewrite the Lagrangian equations of motion in
terms of the mechanical momentum
\begin{equation}
\mathbf{p}_{a}=m_{a}\mathbf{v}_{a}[1+\mathbf{v}_{a}^{2}/(2c^{2})]
\end{equation}
\ Then Newton's second law
\begin{equation}
d\mathbf{p}_{a}/dt=\frac{d}{dt}\{m_{a}\mathbf{v}_{a}[1+\mathbf{v}_{a}%
^{2}/(2c^{2})]\}=e_{a}\mathbf{E(r}_{a},t)\mathbf{+}e_{a}(\mathbf{v}%
_{a}/c)\times\mathbf{B(r}_{a},t)
\end{equation}
is obtained by carrying out the time derivative in the Darwin equations of
motion (3)\ and recognizing the electric field as\cite{P-A}%
\begin{equation}
\mathbf{E}(\mathbf{r}_{a},t)=%
{\displaystyle\sum\limits_{b\neq a}}
\left\{  \frac{e_{b}\mathbf{r}_{ab}}{r_{ab}^{3}}\left[  1+\frac{1}{2}%
\frac{\mathbf{v}_{b}^{2}}{c^{2}}-\frac{3}{2}\frac{(\mathbf{v}_{b}%
\cdot\mathbf{r}_{ab})^{2}}{c^{2}r_{ab}^{2}}\right]  -\frac{e_{b}}{2c^{2}%
r_{ab}}\left[  \mathbf{a}_{b}+\frac{\mathbf{r}_{ab}(\mathbf{r}_{ab}%
\cdot\mathbf{a}_{b})}{r_{ab}^{2}}\right]  \right\}
\end{equation}
and the magnetic field as%
\begin{equation}
\mathbf{B}(\mathbf{r}_{a},t)=%
{\displaystyle\sum\limits_{b\neq a}}
\frac{e_{b}}{c}\frac{\mathbf{v}_{b}\times\mathbf{r}_{ab}}{r_{ab}^{3}}%
\end{equation}
where $\mathbf{a}_{b}$ is the acceleration of particle $b$. \ In Eq. (6), the
terms of order $1/c^{2}$ provide the familiar effects of Faraday induction.
\ We can also write the electromagnetic fields in terms of electromagnetic
potentials as%
\begin{equation}
\mathbf{E}(\mathbf{r}_{a},t)=-\nabla_{a}\Phi(\mathbf{r}_{a},t)-\frac{1}%
{c}\frac{\partial}{\partial t}\mathbf{A}(\mathbf{r}_{a},t)\text{ \ \ and
\ \ }\mathbf{B}(\mathbf{r}_{a},t)=\nabla_{a}\times\mathbf{A}(\mathbf{r}_{a},t)
\end{equation}
where\cite{J-2}%
\begin{equation}
\Phi(\mathbf{r}_{a},t)=%
{\displaystyle\sum\limits_{b\neq a}}
\frac{e_{b}}{r_{ab}}\text{ \ \ and \ \ }\mathbf{A}(\mathbf{r}_{a},t)=%
{\displaystyle\sum\limits_{b\neq a}}
\frac{e_{b}}{2cr_{ab}}\left[  \mathbf{v}_{b}+\frac{\mathbf{r}_{ab}%
(\mathbf{r}_{ab}\cdot\mathbf{v}_{b})}{r_{ab}^{2}}\right]
\end{equation}
We recognize from Eq. (2) and Eq. (9) that%
\begin{equation}
\mathbf{p}_{a}^{canonical}=m_{a}\mathbf{v}_{a}[1+\mathbf{v}_{a}^{2}%
/(2c^{2})]+(e_{a}/c)\mathbf{A}(\mathbf{r}_{a},t)
\end{equation}
where $\mathbf{A}(\mathbf{r}_{a},t)$ is the vector potential due to all the
other charges evaluated at the position $\mathbf{r}_{a}$ of the charge
$e_{a}.$

\section{Two-Particle Model for a Magnetic Moment}

Our model for a magnetic moment will consist of two charge particles of
different mass in Coulomb orbit around each other (a classical hydrogen atom).
\ There is no electromagnetic radiation in the Darwin Lagrangian, and thus the
orbiting charges do not lose energy in this $1/c^{2}$ approximation.
\ Furthermore, for our model, we will average over the phases of orbital
motion and also average over the configurations where the both the charges and
the velocities of the charges are reversed in sign. \ In this fashion one
maintains the magnetic moment behavior while averaging out the irrelevant
electrostatic aspects.

In this article, the motion of the magnetic moment charges is considered
extensively. \ Therefore,\ for simplicity of notation (and in contrast to the
notation of Coleman and Van Vleck), the magnetic moment consists of a particle
of charge $e$, small mass $m,$ displacement $\mathbf{r}$, velocity
$\mathbf{v}$, and acceleration $\mathbf{a}$ in orbit around a massive particle
of charge $-e,$ mass $M$ (with $M>>m$), displacement $\mathbf{R}\cong
m\mathbf{r}/M\cong0$, velocity $\mathbf{V}\cong m\mathbf{v}/M$, and
acceleration $d\mathbf{V}/dt\mathbf{.}$ \ Since the mass $M$ is large compared
to $m$, the displacement $\mathbf{R}$, velocity $\mathbf{V}$, and acceleration
$d\mathbf{V}/dt$ are all small compared to $\mathbf{r}$, $\mathbf{v}$, and
$\mathbf{a}$ respectively. \ The distant point charge with which the magnetic
moment interacts has charge $q$, mass $m_{q}$, displacement $\mathbf{r}_{q}$,
velocity $\mathbf{v}_{q},$ and acceleration $d\mathbf{v}_{q}/dt.$\ Then from
equations (4)-(7), our equations of motion for the charge $e$ in orbit, the
massive particle $-e$, and the distant charge $q$ are respectively%
\begin{align}
\frac{d}{dt}\left[  m\mathbf{v}\left(  1+\frac{1}{2}\frac{\mathbf{v}^{2}%
}{c^{2}}\right)  \right]   &  =e\mathbf{E}_{-e}(\mathbf{r},t)+e\mathbf{E}%
_{q}(\mathbf{r},t)+e\frac{\mathbf{v}}{c}\times\mathbf{B}_{q}(\mathbf{r}%
,t)\nonumber\\
&  =-\frac{e^{2}\mathbf{r}}{r^{3}}+\frac{eq\mathbf{r}_{eq}}{r_{eq}^{3}}\left[
1+\frac{1}{2}\frac{\mathbf{v}_{q}^{2}}{c^{2}}-\frac{3}{2}\frac{(\mathbf{v}%
_{q}\cdot\mathbf{r}_{eq})^{2}}{c^{2}r_{eq}^{2}}\right]  +e\frac{\mathbf{v}}%
{c}\times\left(  \frac{q}{c}\frac{\mathbf{v}_{q}\times\mathbf{r}_{eq}}%
{r_{eq}^{3}}\right)
\end{align}%
\begin{align}
\frac{d}{dt}\left(  M\mathbf{V}\right)   &  =-e\mathbf{E}_{e}(0,t)-e\mathbf{E}%
_{q}(0,t)\nonumber\\
&  =\frac{e^{2}\mathbf{r}}{r^{3}}\left(  1+\frac{1}{2}\frac{\mathbf{v}^{2}%
}{c^{2}}-\frac{3}{2}\frac{(\mathbf{v}\cdot\mathbf{r})^{2}}{c^{2}r^{2}}\right)
+\frac{e^{2}}{2c^{2}r}\left(  \mathbf{a}+\frac{(\mathbf{a}\cdot\mathbf{r}%
)\mathbf{r}}{r^{2}}\right) \nonumber\\
&  -\frac{eq\mathbf{r}_{q}}{r_{q}^{3}}\left[  1+\frac{1}{2}\frac
{\mathbf{v}_{q}^{2}}{c^{2}}-\frac{3}{2}\frac{(\mathbf{v}_{q}\cdot
\mathbf{r}_{q})^{2}}{c^{2}r_{q}^{2}}\right]
\end{align}
and%
\begin{align}
\frac{d}{dt}\left[  m_{q}\mathbf{v}_{q}\left(  1+\frac{1}{2}\frac
{\mathbf{v}_{q}^{2}}{c^{2}}\right)  \right]   &  =q\mathbf{E}_{-e}%
(\mathbf{r}_{q},t)+q\mathbf{E}_{e}(\mathbf{r}_{q},t)+q\frac{\mathbf{v}_{q}}%
{c}\times\mathbf{B}_{e}(\mathbf{r}_{q},t)\nonumber\\
&  =q\frac{-e\mathbf{r}_{q}}{r_{q}^{3}}+q\frac{e\mathbf{r}_{qe}}{r_{qe}^{3}%
}\left(  1+\frac{1}{2}\frac{\mathbf{v}^{2}}{c^{2}}-\frac{3}{2}\frac
{(\mathbf{v}\cdot\mathbf{r}_{qe})^{2}}{c^{2}r_{qe}^{2}}\right) \nonumber\\
&  -q\frac{e}{2c^{2}r_{qe}}\left(  \mathbf{a}+\frac{(\mathbf{a}\cdot
\mathbf{r}_{qe})\mathbf{r}_{qe}}{r_{qe}^{2}}\right)  +q\frac{\mathbf{v}_{q}%
}{c}\times\left(  \frac{e}{c}\frac{\mathbf{v}\times\mathbf{r}_{qe}}{r_{qe}%
^{3}}\right)
\end{align}
where $\mathbf{r}_{qe}=\mathbf{r}_{q}-\mathbf{r=-r}_{eq}$, and we have assumed
that $\mathbf{V}^{2}/c^{2}<<1.$

\subsection{Nonrelativistic Interaction}

In order to understand the interaction represented by these equations of
motion (11)-(13), we consider first the nonrelativistic approximation 0-order
in $1/c^{2}$ \ where the equations become%
\begin{equation}
m\mathbf{a=}-\frac{e^{2}\mathbf{r}}{r^{3}}+e\mathbf{E}_{q}^{(0)}(\mathbf{r},t)
\end{equation}%
\begin{equation}
M\frac{d\mathbf{V}}{dt}=\frac{e^{2}\mathbf{r}}{r^{3}}-e\mathbf{E}_{q}%
^{(0)}(0,t)
\end{equation}
and
\begin{equation}
m_{q}\frac{d\mathbf{v}_{q}}{dt}=q\frac{-e\mathbf{r}_{q}}{r_{q}^{3}}%
+q\frac{e\mathbf{r}_{qe}}{r_{qe}^{3}}%
\end{equation}
Here the small electrostatic field of the charge $q$ is essentially uniform
across the magnetic moment
\begin{equation}
\mathbf{E}_{q}^{(0)}(\mathbf{r},t)=\frac{q(\mathbf{r}-\mathbf{r}_{q}%
)}{|\mathbf{r}-\mathbf{r}_{q}|^{3}}\cong-\frac{q\mathbf{r}_{q}}{r_{q}^{3}%
}=\mathbf{E}_{q}^{(0)}(0,t)
\end{equation}
since the charge $q$ is distant from the magnetic moment at the origin of
coordinates, $r/r_{q}<<1.~$\ The electrostatic field at the charge $q$
appearing on the right-hand side in Eq. (16) is an electric dipole field and
is even smaller (for $q$ and $e$ of the same magnitude) because the magnetic
moment is electrically neutral. \ 

In this nonrelativistic approximation, the interaction of the distant point
charge $q$ with this magnetic moment depends crucially upon the orientation of
the magnetic moment. \ i)If the magnetic moment $\overrightarrow{\mu}$ at the
origin is aligned parallel to the displacement $\mathbf{r}_{q}$\ to the point
charge, $\overrightarrow{\mu}||\mathbf{r}_{q}$, we find the stable
electrostatic polarizability aspect. \ ii)If the magnetic moment
$\overrightarrow{\mu}$\ is aligned perpendicular to the displacement
$\mathbf{r}_{q}$ to the point charge, $\overrightarrow{\mu}\bot\mathbf{r}_{q}%
$, then we find Solem's\cite{Solem} unstable "strange polarizability" aspect.
\ It is the second, unfamiliar aspect which is crucial for understanding the
electric forces which are proportional to the magnetic moment. \ 

\subsubsection{Stable Electrostatic Polarizability}

If the distant charge $q$ lies along the axis perpendicular to the orbital
motion and through its center, $\overrightarrow{\mu}||\mathbf{r}_{q}$, then
the electric field $\mathbf{E}_{q}^{(0)}$ will cause a displacement $l$\ of
the orbital plane relative to the massive particle $M.$ \ The equilibrium
situation for the orbital motion with angular frequency $\omega$ corresponds
to Newton's equations of motion in the radial and axial directions giving
\begin{equation}
m\omega^{2}r=e^{2}r\,(r^{2}+l^{2})^{-3/2}\text{ \ \ \ \ and \ \ \ \ }%
eE_{q}^{(0)}=e^{2}l\,(r^{2}+l^{2})^{-3/2}%
\end{equation}
Eliminating $r$ between the equations, we find $e^{2}/(m\omega^{2})E_{q}%
^{(0)}=el=\mathfrak{p},$ where $\mathfrak{p}$ is the average electric dipole
moment of the two-particle magnetic moment. \ Thus the magnetic moment in this
orientation has an electrostatic polarizability%

\begin{equation}
\alpha=e^{2}/(m\omega^{2})\text{ \ \ where \ \ }\overrightarrow{\mathfrak{p}%
}=\alpha\mathbf{E}_{q}^{(0)}%
\end{equation}
a form for $\alpha$ which is familiar for a dipole harmonic
oscillator.\cite{J3} \ We notice that the polarizability is even in the charge
$e$ and in the frequency $\omega$ and has no relation to the sign of the
magnetic moment
\begin{equation}
\overrightarrow{\mu}=e\mathbf{L}/((2mc)=e\overrightarrow{\omega}r^{2}/(2c)
\end{equation}

\subsubsection{Solem's Unstable "Strange" Polarization}

If the magnetic moment is oriented perpendicular to the displacement to the
distant charge $q,$ $\overrightarrow{\mu}\bot\mathbf{r}_{q}$, then we find
behavior which is mentioned only rarely in the physics literature.\cite{Solem}
\ It does not appear in Coleman and Van Vleck's article,\cite{CVV} but it is
crucial to understanding the classical electromagnetic interactions associated
with the Aharonov-Bohm and Aharonov-Casher phase shifts. \ In this case when
the angular momentum $\mathbf{L}$ of the orbit for the magnetic moment is
perpendicular to the electric field $\mathbf{E}_{q}^{(0)}$ of the distant
charge $q$,$\overrightarrow{\mu}\bot\mathbf{r}_{q},$ the initial circular
orbit is transformed into an elliptical orbit of ever-changing ellipticity
with its semi-major axis oriented perpendicular to both the angular momentum
$\mathbf{L}$ and the electric field $\mathbf{E}_{q}^{(0)}$.\cite{Solem} \ In
order to analyze this motion, it is useful to introduce the Laplace-Runge-Lenz
vector $\mathbf{K}$ for the Coulomb orbit of the charge $e.$\cite{Gold} \ We
assume that the much larger mass $M$ is at the origin, $\mathbf{R}\cong0,$ so
that the charge $e$ moves with a displacement\cite{Solem}
\begin{equation}
\mathbf{r}=\frac{3}{2}\frac{\mathbf{K}}{(-2mH_{0})^{1/2}}+\frac{1}{4H_{0}%
}\frac{d}{dt}[m(\mathbf{r}\times\mathbf{v})\times\mathbf{r}+m\mathbf{v}r^{2}]
\end{equation}
where $\mathbf{K}$ is the Laplace-Runge-Lenz vector\cite{Gold}%
\begin{equation}
\mathbf{K}=\frac{1}{(-2mH_{0})^{1/2}}\left(  [\mathbf{r}\times
(m\mathbf{v)]\times(}m\mathbf{v)+}me^{2}\frac{\mathbf{r}}{r}\right)
\end{equation}
and $H_{0}$ is the particle energy
\begin{equation}
H_{0}=mv^{2}/2-e^{2}/r
\end{equation}
The equation (21) can be checked by carrying out the time derivative and then
inserting the equation of motion $\mathbf{a}=-e^{2}\mathbf{r}/(mr^{3})$ for
every appearance of the acceleration $\mathbf{a=}d\mathbf{v}/dt=d^{2}%
\mathbf{r}/dt^{2}.$ \ The Laplace-Runge-Lenz vector is constant in time for a
Coulomb orbit, and the second term of (21) involving a time derivative shows
how the displacement $\mathbf{r}$ varies in time. \ On time-averaging, the
time derivative vanishes leaving%
\begin{equation}
<\mathbf{r}>=\frac{3}{2}\frac{\mathbf{K}}{(-2mH_{0})^{1/2}}%
\end{equation}
The average electric dipole moment $\overrightarrow{\mathfrak{p}}$ is given
by
\begin{equation}
\overrightarrow{\mathfrak{p}}=e<\mathbf{r}>=\frac{3}{2}\frac{e\mathbf{K}%
}{(-2mH_{0})^{1/2}}%
\end{equation}
We assume that initially the magnetic moment has a circular orbit for the
charge $e$, and therefore initially the electric dipole moment vanishes,
$\overrightarrow{\mathfrak{p}}=e<\mathbf{r}>=0$ and $\mathbf{K}=0.$ \ However,
in the presence of the electric field $\mathbf{E}_{q}^{(0)}$ of the distant
charge $q$, the equation of motion for $e$ is given in Eq. (14). \ We assume
that the field $\mathbf{E}_{q}^{(0)}$ is small so that the orbit remains
Coulombic but now with a slowly changing Laplace-Runge-Lenz vector. \ The time
rate of change of $\mathbf{K}$ can be obtained by differentiating both sides
of equation (22) and the use of the equation of motion (14),%
\begin{align}
\frac{d\mathbf{K}}{dt}  &  =\frac{1}{(-2mH_{0})^{1/2}}m\left\{  \left[
\mathbf{r}\times\left(  \frac{-e^{2}\mathbf{r}}{r^{3}}+e\mathbf{E}_{q}\right)
\right]  \times\mathbf{v}+(\mathbf{r}\times\mathbf{v})\times\left(
\frac{-e^{2}\mathbf{r}}{r^{3}}+e\mathbf{E}_{q}\right)  +e^{2}\left[
\frac{\mathbf{v}}{r}-\frac{\mathbf{r}(\mathbf{r}\cdot\mathbf{v})}{r^{3}%
}\right]  \right\} \nonumber\\
&  =(-2mH_{0})^{-1/2}me[-2\mathbf{r}(\mathbf{v\cdot E}_{q}^{(0)}%
)+\mathbf{E}_{q}^{(0)}(\mathbf{r}\cdot\mathbf{v})+\mathbf{v}(\mathbf{r}%
\cdot\mathbf{E}_{q}^{(0)})]
\end{align}
We note again that the Laplace-Runge-Lenz vector would be constant in time
were it not for the external electric field $\mathbf{E}_{q}^{(0)}.$ \ Since we
assume that the distant charge $q$ is causing a small perturbation, we may
average the particle displacement $\mathbf{r}$ and velocity $\mathbf{v}$ over
an orbit of the unperturbed motion. \ Now if $f(\mathbf{r,v)}$ is any function
of the displacement and velocity of the unperturbed orbit, then it is a
periodic function in time with period given by the orbital period $T$.
\ Therefore, the time average of the time derivative vanishes%
\[
\left\langle \frac{d}{dt}f(\mathbf{r,v)}\right\rangle =\frac{1}{T}%
{\displaystyle\int\limits_{0}^{t=T}}
dt\,\frac{d}{dt}f(\mathbf{r,v)}=0
\]
In particular for $f(\mathbf{r,v)}=x_{i}x_{j}$ where $x_{i}~$and $x_{j}$ are
the $i$th and $j$th components of $\mathbf{r}$, then we have%
\begin{equation}
\left\langle \frac{d}{dt}(x_{i}x_{j})\right\rangle =\left\langle x_{i}%
v_{j}\right\rangle +\left\langle x_{j}v_{i}\right\rangle =0
\end{equation}
so that%
\begin{align}
\left\langle \mathbf{r}\cdot\mathbf{v}\right\rangle  &  =0\nonumber\\
\left\langle \mathbf{r}\left(  \mathbf{v}\cdot\mathbf{E}_{q}^{(0)}\right)
\right\rangle  &  =\left\langle -\mathbf{v}\left(  \mathbf{r}\cdot
\mathbf{E}_{q}^{(0)}\right)  \right\rangle =-\left\langle \left(
\mathbf{r}\times\mathbf{v}\right)  \times\mathbf{E}_{q}^{(0)}\right\rangle /2
\end{align}
This result allows us to average over the unperturbed motion to obtain%

\begin{align}
(-2mH_{0})^{1/2}d\mathbf{K/}dt  &  =\left\langle me[-2\mathbf{r}%
(\mathbf{v\cdot E}_{q}^{(0)})+\mathbf{E}_{q}^{(0)}(\mathbf{r}\cdot
\mathbf{v})+\mathbf{v}(\mathbf{r}\cdot\mathbf{E}_{q}^{(0)})]\right\rangle
\nonumber\\
&  =(3/2)me[\left\langle \mathbf{r}\times\mathbf{v}\right\rangle
\times\mathbf{E}_{q}^{(0)}]=(3/2)e\mathbf{L}\times\mathbf{E}_{q}%
^{(0)}=3m(c\overrightarrow{\mu})\times\mathbf{E}_{q}^{(0)}%
\end{align}
where $\mathbf{L}$ is the angular momentum of the orbit and $\overrightarrow
{\mu}=e\mathbf{L}/(2mc)$. \ Thus from Eqs. (25) and (29), the electric dipole
moment is changing as%
\begin{equation}
\frac{d\overrightarrow{\mathfrak{p}}}{dt}=\frac{9}{4}\frac{e^{2}}{(-2mH_{0}%
)}\mathbf{L}\times\mathbf{E}_{q}^{(0)}%
\end{equation}

This is a very strange polarization indeed. \ The initially unpolarized orbit
does indeed develop an electrical polarization with time, but the\ predominant
electric dipole moment depends upon the orbital angular momentum and is in a
direction perpendicular to the applied electric field $\mathbf{E}_{q}^{(0)}$.
\ Since the angular momentum $\mathbf{L}$ is related to the magnetic moment as
$\overrightarrow{\mu}=e\mathbf{L}/(2mc)$, we have the developing polarization
related to the magnetic moment. \ However, if we average over both the orbital
positions and over both signs $\pm e$ of charge while maintaining the
direction of the magnetic moment $\overrightarrow{\mu}=e\overrightarrow
{\omega}r^{2}/(2c),$ then we see that the time rate of change of the
Laplace-Runge-Lenz vector $\mathbf{K}$ does not average to zero while the
average rate of change of polarization $\left\langle d\overrightarrow
{\mathfrak{p}}/dt\right\rangle $ actually vanishes, since the direction of
angular momentum in Eq. (20) reverses as the sign of the charge $e$ is
reversed. \ We also notice that the rate of change of the Laplace-Runge-Lenz
vector and of the electric dipole moment for an individual orbit depends upon
the value of the field $\mathbf{E}_{q}^{(0)}$ alone and is independent of any
rate of change of the electric field $\mathbf{E}_{q}^{(0)}$. \ This is
completely different from the electrical polarization $\overrightarrow
{\mathfrak{p}}$\ found from the electrostatic polarizability in Eq. (19) where
there is no \textit{change} in the polarization unless the field
$\mathbf{E}_{q}^{(0)}$ changes in time.

There are additional observations which should be made regarding the behavior
of the magnetic moment under the action of the electric field $\mathbf{E}%
_{q}^{(0)}$ of the distant point charge $q$. \ The sum of the particle kinetic
energy plus electrostatic potential energy is conserved. \ Indeed, while the
average displacement $<\mathbf{r}>$ of the charge $e$ is initially zero and
increases in time, the length of the semimajor axis of the orbit does not
change and is oriented in a direction perpendicular to the electric field
$\mathbf{E}_{q}^{(0)}$; the work done by the electric field $\mathbf{E}%
_{q}^{(0)}$ on the orbiting charge $e$ vanishes when averaged over the Coulomb
orbit. \ The average position of the heavier mass $M$ with charge $-e$ also
shifts slightly so as to maintain the position of the center of (rest) mass of
the magnetic moment system at the origin; since the average electrostatic
force on the magnetic moment (due to the uniform electric field $\mathbf{E}%
_{q}^{(0)}$ of the point charge $q$) vanishes, the position of the center of
(rest) mass does not change. \ As the orbiting system develops an electric
dipole moment $\overrightarrow{\mathfrak{p}}$, there are balancing
electrostatic forces and torques on the magnetic moment due to the point
charge and on the point charge due to the magnetic moment. \ However, when we
average over magnetic moments carrying opposite charges $\pm e$ but the same
magnetic moment $\overrightarrow{\mu}=e\overrightarrow{\omega}r^{2}/(2c)$, all
of the dipole-associated electrostatic forces and torques vanish in the average.

\subsection{Electromagnetic Forces on the Distant Point Charge}

\subsubsection{Force Associated with the Stable Electrostatic Polarization}

Having obtained the behavior of the magnetic moment in the 0-order
nonrelativistic system, we now wish to consider the electromagnetic forces
$\mathbf{F}_{on\,q}\mathbf{=}q\mathbf{E}_{\mu}+q(\mathbf{v}_{q}/c)\times
\mathbf{B}_{\mu}$ acting on the distant point charge $q$\ due to the magnetic
moment $\overrightarrow{\mu}$. \ \ The forces are different depending upon the
orientation of the magnetic moment. \ When the magnetic moment
$\overrightarrow{\mu}$ is parallel to the displacement $\mathbf{r}_{q}$ to the
distant charge $q,$ $\overrightarrow{\mu}||\mathbf{r}_{q},$ then we saw in Eq.
(19) that the magnetic moment has an induced electric dipole moment
$\overrightarrow{\mathfrak{p}}=\alpha\mathbf{E}_{q}.$\ Accordingly, the
electrically polarized magnetic moment creates an electrostatic dipole field
$\mathbf{E}_{\mathfrak{p}}(\mathbf{r}_{q},t)$ which causes an electrostatic
force $\mathbf{F}_{on\text{ }q}$ on $q$ given by%
\begin{equation}
\mathbf{F}_{on\text{ }q}=q\mathbf{E}_{\mathfrak{p}}(\mathbf{r}_{q}%
,t)=q\{2\overrightarrow{\mathfrak{p}}\}r_{q}^{-3}=q\{2[e^{2}/(m\omega
^{2})]\mathbf{E}_{q}(0,t)\}r_{q}^{-3}=-\mathbf{r}_{q}q^{2}e^{2}/(m\omega
^{2}r_{q}^{7})
\end{equation}
The electrostatic force back at the charge $q$ is independent of the sign of
the charge $q,$ or of the sign of the charge $e,$ or of the direction of
rotation $\omega$. \ When averaged over the orbital motion and over both signs
of charge $\pm e$ for the magnetic moment, the only force on $q$ is this
electrostatic dipole force. \ There is no additional force of order $1/c^{2}$.
\ As an aside, we note that for this orientation of the magnetic moment,
$\overrightarrow{\mu}||\mathbf{r}_{q}$, the magnetic vector potential
$\mathbf{A}_{\mu}$ vanishes along the axis through the magnetic moment
parallel to the magnetic moment direction.

\subsubsection{Force Associated with Solem's Unstable "Strange" Polarization}

The situation is completely different when the magnetic moment is oriented
perpendicular to the displacement $\mathbf{r}_{q},$ $\overrightarrow{\mu}%
\bot\mathbf{r}_{q}$. \ In this case we saw that after carrying out the
averaging for the magnetic moment, there were no electric monopole or dipole
contributions to a force back on the point charge $q$. \ Since these 0-order
back forces vanish, the back forces in order $1/c^{2}$ caused by the 0-order
changes of the magnetic moment are of considerable interest. \ The alteration
in the shape of the Coulomb orbit leads to unbalanced accelerations
$\mathbf{a}$ which lead to new contributions to the electric field according
to Eq. (6). \ The vector potential in the Coulomb gauge of a point charge $e$
is given in Eq. (9), \ and we see that the last term in Eq. (6) corresponds to
the electric field contribution from $-\partial\left\langle \mathbf{A}%
_{e}\right\rangle /\partial t=-\partial\mathbf{A}_{\mu}/\partial t$. \ Now the
magnetic moment model corresponds to a magnetic moment given initially by
$\overrightarrow{\mu}=e\overrightarrow{\omega}r^{2}/(2c)$ in Eq. (20), and
therefore to a vector potential
\begin{equation}
\mathbf{A}_{\mu}(\mathbf{r},t)=\frac{\overrightarrow{\mu}\times\mathbf{r}%
}{cr^{3}}=\frac{e}{2mc^{2}}\frac{\mathbf{L}\times\mathbf{r}}{r^{3}}%
\end{equation}
Thus for our magnetic moment model, the average electric field $\mathbf{E}%
_{\mu}$ back at the charged particle $q$ will be related to the change in the
angular momentum $\mathbf{L}$ of the orbit. \ Now the change in angular
momentum $\mathbf{L}$ of the orbit of the charge $e$ is due solely to the
presence of the external charge $q$ which gives $d\mathbf{L}/dt=\mathbf{r}%
\times e\mathbf{E}_{q}^{(0)}$, and, when averaged over one period of the
motion, becomes from Eq. (24)%
\begin{equation}
\frac{d\mathbf{L}}{dt}=\left\langle \mathbf{r}\right\rangle \times
e\mathbf{E}_{q}^{(0)}=-e\mathbf{E}_{q}^{(0)}\times\left\langle \mathbf{r}%
\right\rangle =-e\mathbf{E}_{q}^{(0)}\times\frac{3}{2}\frac{\mathbf{K}%
}{(-2mH_{0})^{1/2}}%
\end{equation}
Thus the electric field back at the charge $q$ is given by%
\begin{align}
\mathbf{E}_{\mu}\mathbf{(r}_{q},t)  &  =-\frac{1}{c}\frac{\partial}{\partial
t}\mathbf{A}_{\mu}(\mathbf{r}_{q},t)=\frac{e}{2mc}\frac{\mathbf{r}_{q}}%
{r_{q}^{3}}\times\frac{d\mathbf{L}}{dt}\nonumber\\
&  =\frac{e}{2mc^{2}}\frac{\mathbf{r}_{q}}{r_{q}^{3}}\times\left(
-e\mathbf{E}_{q}^{(0)}\times\frac{3}{2}\frac{\mathbf{K}}{(-2mH_{0})^{1/2}%
}\right)
\end{align}
Now our magnetic moment model is initially in a circular orbit with
$\mathbf{K}=0,$ and $\mathbf{K}$\ changed \ as in Eq. (29) only because of the
presence of the electric field $\mathbf{E}_{q}^{(0)}$ due to the distant
charge $q.$ \ Thus the force $\mathbf{F}_{on\text{ }q}$ on $q$ due to the
electric field $\mathbf{E}_{\mu}$ of the magnetic moment is
\begin{equation}
\mathbf{F}_{on\text{ }q}=q\mathbf{E}_{\mu}\mathbf{(r}_{q},t)=q\frac{e}%
{2mc^{2}}\frac{\mathbf{r}_{q}}{r_{q}^{3}}\times\left(  \frac{3}{2}%
\frac{-e\mathbf{E}_{q}^{(0)}(0,t^{\prime})}{(-2mH_{0})^{1/2}}\times%
{\displaystyle\int\limits_{0}^{t}}
dt^{\prime}\,\{3m[c\overrightarrow{\mu}(t^{\prime})]\times\mathbf{E}_{q}%
^{(0)}(0,t^{\prime})\}\right)
\end{equation}
where $\mathbf{E}_{q}^{(0)}$ is the electrostatic field in Eq. (17) of the
distant charge $q$ acting on the magnetic moment. \ We notice that this force
back on the charge $q$ due to the magnetic moment $\overrightarrow{\mu}$ is
proportional to $q^{3}e^{2}\mu$; it changes sign with the external charge $q$,
changes sign with the magnetic moment $\overrightarrow{\mu},$ but does not
depend upon the sign of the charge $e$. \ Furthermore it changes sign with the
reversal of the position $\mathbf{r}_{q}$ of the charge $q$. \ Finally, it
does not depend upon any velocity of the charge $q$. It arises from the
0-order acceleration of the orbiting magnetic moment charge due to the
\textit{electrostatic}\ field $\mathbf{E}_{q}^{(0)}$ of the distant charge
$q.$ \ These properties are in total contrast with those found for
electrostatic forces such as in Eq. (31).

\section{Conservation Laws}

In our model, the (zero-order) electrostatic field of the passing charge
causes a change in the magnetic moment which then produces an (order $1/c^{2}%
$) electric field back at the position of the passing charge. \ Since this
back electric field is unanticipated by treatments (such as in the
no-velocity-change point of view) which do not allow for changes in the charge
and current densities of magnetic moments, it seems appropriate to discuss all
the conservation laws associated with electromagnetic theory and to see how
they are upheld by the present model.

\subsection{Linear Momentum in the Electromagnetic Field}

The Darwin Lagrangian conserves linear momentum\cite{Momentum}. \ For our
magnetic moment and passing charge, the total linear momentum is%
\begin{align}
\mathbf{P}  &  =\mathbf{P}_{\mu}+\mathbf{P}_{em\;\mu q}+\mathbf{p}%
_{q}\nonumber\\
&  =\left[  M\mathbf{V+}m\mathbf{v}\left(  1+\frac{1}{2}\frac{\mathbf{v}^{2}%
}{c^{2}}\right)  -\frac{e^{2}}{2c^{2}r}\left(  \mathbf{v}+\frac{[\mathbf{v}%
\cdot\mathbf{r}]\mathbf{r}}{r^{2}}\right)  \right] \nonumber\\
&  +\left[  \frac{qe}{2c^{2}|\mathbf{r}_{q}-\mathbf{r}|}\left(  \mathbf{v}%
+\frac{[\mathbf{v}\cdot(\mathbf{r}_{q}\mathbf{-r})](\mathbf{r}_{q}%
\mathbf{-r})}{|\mathbf{r}_{q}\mathbf{-r}|^{2}}\right)  \right]  +\left[
m_{q}\mathbf{v}_{q}\left(  1+\frac{1}{2}\frac{\mathbf{v}_{q}^{2}}{c^{2}%
}\right)  \right]
\end{align}
Here we have grouped the total momentum into three terms which can be assigned
to the magnetic moment, the electromagnetic fields between the magnetic moment
and the charge $q$, and the mechanical momentum of the passing charge $q$.
\ When averaged over the orbital motion of the magnetic moment, the system
carries an average linear momentum in the electromagnetic field given by%
\begin{align}
\left\langle \mathbf{P}_{em\,\mu q}\right\rangle  &  =\left\langle \frac
{1}{4\pi c}%
{\displaystyle\int}
d^{3}r\,\mathbf{E}_{q}\times\mathbf{B}_{\mu}\right\rangle \nonumber\\
&  =\left\langle \frac{qe}{2c^{2}|\mathbf{r}_{q}-\mathbf{r}|}\left(
\mathbf{v}+\frac{[\mathbf{v}\cdot(\mathbf{r}_{q}\mathbf{-r})](\mathbf{r}%
_{q}\mathbf{-r})}{|\mathbf{r}_{q}\mathbf{-r}|^{2}}\right)  \right\rangle
\nonumber\\
&  =\frac{q}{c}\frac{\overrightarrow{\mu}\times\mathbf{r}_{q}}{r_{q}^{3}%
}=\frac{q}{c}\mathbf{A}_{\mu}(\mathbf{r}_{q},t)
\end{align}
where, from Eq. (9), $\mathbf{A}_{\mu}(\mathbf{r}_{q},t)$ is the vector
potential in the Coulomb gauge due to the magnetic moment and evaluated at the
position of the point charge $q$. \ Any contribution from the other
electromagnetic field combination $\mathbf{E}_{\mu}\times\mathbf{B}_{q}$ is
very small since the magnetic moment $\overrightarrow{\mu}$ has no net charge. \ 

Now the time derivative of the electromagnetic field momentum $\left\langle
\mathbf{P}_{em\,\mu q}\right\rangle $ in Eq. (37) involves changes connected
with the particle position $\mathbf{r}_{q}$ and with the magnetic moment
$\overrightarrow{\mu}$. \ We can write%
\begin{align}
\frac{d}{dt}\left\langle \mathbf{P}_{em\,\mu q}\right\rangle  &  =\frac{d}%
{dt}\left(  \frac{q}{c}\mathbf{A}_{\mu}(\mathbf{r}_{q},t)\right) \nonumber\\
&  =(\mathbf{v}_{q}\cdot\nabla_{q})\left(  \frac{q}{c}\mathbf{A}_{\mu
}(\mathbf{r}_{q},t)\right)  +\frac{\partial}{\partial t}\left(  \frac{q}%
{c}\mathbf{A}_{\mu}(\mathbf{r}_{q},t)\right) \nonumber\\
&  =\nabla_{q}\left(  \frac{q}{c}\mathbf{v}_{q}\cdot\mathbf{A}_{\mu
}(\mathbf{r}_{q},t)\right)  -\frac{q}{c}\mathbf{v}_{q}\times\left[  \nabla
_{q}\times\mathbf{A}_{\mu}(\mathbf{r}_{q},t)\right]  +\frac{\partial}{\partial
t}\left(  \frac{q}{c}\mathbf{A}_{\mu}(\mathbf{r}_{q},t)\right) \nonumber\\
&  =\nabla_{q}\left(  \frac{q}{c}\mathbf{v}_{q}\cdot\frac{\overrightarrow{\mu
}\times\mathbf{r}_{q}}{r_{q}^{3}}\right)  -\frac{q}{c}\mathbf{v}_{q}%
\times\left[  \nabla_{q}\times\left(  \frac{\overrightarrow{\mu}%
\times\mathbf{r}_{q}}{r_{q}^{3}}\right)  \right]  +\frac{q}{c}\left(
\frac{d\overrightarrow{\mu}}{dt}\times\frac{\mathbf{r}_{q}}{r_{q}^{3}}\right)
\nonumber\\
&  =-\left\langle \mathbf{F}_{on\,\mu}^{Lorentz}\right\rangle -\left\langle
\mathbf{F}_{on\,q}^{Lorentz}\right\rangle
\end{align}
where%
\begin{equation}
\left\langle \mathbf{F}_{on\,\mu}^{Lorentz}\right\rangle =\nabla_{\mathbf{r}%
}\left[  \overrightarrow{\mu}\cdot\mathbf{B}_{q}(\mathbf{r},t)\right]
_{r=0}=-\nabla_{q}\left(  \frac{q}{c}\mathbf{v}_{q}\cdot\mathbf{A}_{\mu
}(\mathbf{r}_{q},t)\right)
\end{equation}
and
\begin{equation}
\left\langle \mathbf{F}_{on\,q}^{Lorentz}\right\rangle =q\mathbf{E}_{\mu
}(\mathbf{r}_{q},t)+q\frac{\mathbf{v}_{q}}{c}\times\mathbf{B}_{\mu}%
(\mathbf{r}_{q},t)
\end{equation}
with
\begin{equation}
\mathbf{E}_{\mu}(\mathbf{r}_{q},t)=-\frac{\partial}{\partial t}\mathbf{A}%
_{\mu}(\mathbf{r}_{q},t)=-\frac{d\overrightarrow{\mu}}{dt}\times
\frac{\mathbf{r}_{q}}{r_{q}^{3}}%
\end{equation}
and%
\begin{equation}
\mathbf{B}_{\mu}(\mathbf{r}_{q},t)=\nabla_{q}\times\mathbf{A}_{\mu}%
(\mathbf{r}_{q},t)=\nabla_{q}\times\left(  \frac{q}{c}\frac{\overrightarrow
{\mu}\times\mathbf{r}_{q}}{r_{q}^{3}}\right)
\end{equation}
Thus the average electromagnetic linear momentum $\left\langle \mathbf{P}%
_{em\;\mu q}\right\rangle $ in Eq. (37) changes with respect to time for two
reasons: the change in $\overrightarrow{\mu}$ (due to the change in the
orbital shape of the magnetic moment) and the change in the separation
$\mathbf{r}_{q}$. \ As the shape changes for the orbit of the charge $e$ in
the magnetic moment, the magnetic moment $\overrightarrow{\mu}$ changes
creating an \textit{electric} field at the position of the passing particle
$q$. \ Thus due to this changing-$\mu$ effect, the linear momentum
$\left\langle \mathbf{P}_{em\;\mu q}\right\rangle $ in the electromagnetic
field decreases at the same rate that the linear momentum of the point charge
$q$ increases due to the force from the electric field of the changing
magnetic moment. \ The change in the electromagnetic linear momentum
$\left\langle \mathbf{P}_{em\;\mu q}\right\rangle $ due to the changing
position $\mathbf{r}_{q}$ is associated with the magnetic Lorentz forces on
the magnetic moment and on the passing charge.

Next we average the total system momentum in Eq. (36) over the orbital motion
and differentiate with respect to time to find%
\begin{align}
\frac{d\mathbf{P}}{dt}  &  =0=\left[  \frac{d\left\langle \mathbf{P}_{\mu
}\right\rangle }{dt}+\nabla_{q}\left(  \frac{q}{c}\frac{\overrightarrow{\mu
}\times\mathbf{r}_{q}}{r_{q}^{3}}\right)  \right] \nonumber\\
&  +\left[  \frac{q}{c}\left(  \frac{d\overrightarrow{\mu}}{dt}\right)
\times\frac{\mathbf{r}_{q}}{r_{q}^{3}}-\frac{q}{c}\mathbf{v}_{q}\times\left[
\nabla_{q}\times\left(  \frac{q}{c}\frac{\overrightarrow{\mu}\times
\mathbf{r}_{q}}{r_{q}^{3}}\right)  \right]  +\frac{d\mathbf{p}_{q}}{dt}\right]
\nonumber\\
&  =\left[  \frac{d\left\langle \mathbf{P}_{\mu}\right\rangle }{dt}%
-\left\langle \mathbf{F}_{on\,\mu}^{Lorentz}\right\rangle \right]  +\left[
\frac{d\mathbf{p}_{q}}{dt}-\left\langle \mathbf{F}_{on\,q}^{Lorentz}%
\right\rangle \right]
\end{align}
\ The equations of motion tell us that each of the quantities in square
brackets vanishes. \ Note that the sum of the average Lorentz forces
$\left\langle \mathbf{F}_{on\,\mu}^{Lorentz}\right\rangle +\left\langle
\mathbf{F}_{on\,q}^{Lorentz}\right\rangle $ does not vanish, but rather
(according to Eq. (38)) is equal to the negative rate of change of the
electromagnetic field linear momentum $\left\langle \mathbf{P}_{em\,\mu
q}\right\rangle $. \ Thus in the conservation law for linear momentum, the
changing electromagnetic field momentum $\left\langle \mathbf{P}_{em\,\mu
q}\right\rangle $ is partially balanced by the changing momentum of the
magnetic moment and partially balanced by the changing momentum of the passing particle.

\subsection{Energy Conservation}

The Darwin lagrangian conserves energy.\cite{energy} \ \ For our magnetic
moment and passing charge, the total energy through order $1/c^{2}$ is%
\begin{align}
U  &  =U_{\mu}+U_{em\;\mu q}+U_{q}\nonumber\\
&  =\left[  Mc^{2}+mc^{2}\left(  1+\frac{1}{2}\frac{v^{2}}{c^{2}}+\frac{3}%
{8}\frac{v^{4}}{c^{4}}\right)  -\frac{e^{2}}{r}\right]  +[-\frac{eq}{r_{q}%
}+\frac{eq}{r_{eq}}\nonumber\\
&  +\frac{eq}{2c^{2}r_{eq}}\left(  \mathbf{v\cdot v}_{q}+\frac{(\mathbf{v}%
\cdot\mathbf{r}_{eq})(\mathbf{v}_{q}\cdot\mathbf{r}_{eq})}{r_{eq}^{2}}\right)
]+\left[  m_{q}c^{2}\left(  1+\frac{1}{2}\frac{v_{q}^{2}}{c^{2}}+\frac{3}%
{8}\frac{v_{q}^{4}}{c^{4}}\right)  \right]
\end{align}
When averaged over the orbital motion of the magnetic moment, the
electrostatic energy $-eq/r_{q}+eq/r_{eq}$ involves only quadrupole energies,
which vanish when averaged $\pm e,$ $\pm\omega$ with $\overrightarrow{\mu}$
held constant. \ The system carries an average magnetic energy in the
electromagnetic field given by
\begin{align}
\left\langle U_{em\;\mu q}\right\rangle  &  =\left\langle \frac{1}{8\pi}%
{\displaystyle\int}
d^{3}r\,\mathbf{B}_{q}\times\mathbf{B}_{\mu}\right\rangle \nonumber\\
&  =\left\langle \frac{eq}{2c^{2}r_{eq}}\left(  \mathbf{v\cdot v}_{q}%
+\frac{(\mathbf{v}\cdot\mathbf{r}_{eq})(\mathbf{v}_{q}\cdot\mathbf{r}_{eq}%
)}{r_{eq}^{2}}\right)  \right\rangle \nonumber\\
&  =\overrightarrow{\mu}\cdot\mathbf{B}_{q}(0,t)=\overrightarrow{\mu}%
\cdot\left(  \frac{q}{c}\frac{\mathbf{v}_{q}\times(-\mathbf{r}_{q})}{r_{q}%
^{3}}\right) \nonumber\\
&  =\frac{q}{c}\mathbf{v}_{q}\cdot\frac{\overrightarrow{\mu}\times
\mathbf{r}_{q}}{r_{q}^{3}}=\frac{q}{c}\mathbf{v}_{q}\cdot\mathbf{A}_{\mu
}(\mathbf{r}_{q},t)
\end{align}
The time derivative of the magnetic field energy $\left\langle U_{em\;\mu
q}\right\rangle $ can be written using the calculations in Eq. (38) for
$d\mathbf{A}_{\mu}/dt$%
\begin{align}
\frac{d}{dt}\left\langle U_{em\;\mu q}\right\rangle  &  =\frac{d}{dt}\left(
\frac{q}{c}\mathbf{v}_{q}\cdot\mathbf{A}_{\mu}(\mathbf{r}_{q},t)\right)
=\mathbf{v}_{q}\cdot\frac{d}{dt}\left(  \frac{q}{c}\mathbf{A}_{\mu}%
(\mathbf{r}_{q},t)\right) \nonumber\\
&  =\mathbf{v}_{q}\cdot\left(  -\left\langle \mathbf{F}_{on\,\mu}%
^{Lorentz}\right\rangle -\left\langle \mathbf{F}_{on\,q}^{Lorentz}%
\right\rangle \right)
\end{align}
since $\left\langle U_{em\;\mu q}\right\rangle $ is already of order $1/c^{2}$
and any change in $\mathbf{v}_{q}$ due to the changing magnetic moment
$\overrightarrow{\mu}$ is also of order $1/c^{2}.$ \ 

Since the total energy in Eq. (44) is constant in time, it follows from
averaging over the orbital motion and differentiating with respect to time
that
\begin{align}
\frac{dU}{dt} &  =0=\frac{d\left\langle U_{\mu}\right\rangle }{dt}+\frac
{d}{dt}\left\langle U_{em\;\mu q}\right\rangle +\frac{dU_{q}}{dt}\nonumber\\
&  =\left[  \frac{d\left\langle U_{\mu}\right\rangle }{dt}+\frac{q}%
{c}\mathbf{v}_{q}\cdot\left(  \overrightarrow{\mu}\times\frac{d}{dt}%
\frac{\mathbf{r}_{q}}{r_{q}^{3}}\right)  \right]  +\left[  \frac{q}%
{c}\mathbf{v}_{q}\cdot\left(  \frac{d\overrightarrow{\mu}}{dt}\times
\frac{\mathbf{r}_{q}}{r_{q}^{3}}\right)  +\frac{dU_{q}}{dt}\right]
\nonumber\\
&  =\left[  \frac{d\left\langle U_{\mu}\right\rangle }{dt}-\mathbf{v}_{q}%
\cdot\left\langle \mathbf{F}_{on\,\mu}^{Lorentz}\right\rangle \right]
+\left[  \frac{dU_{q}}{dt}-\mathbf{v}_{q}\cdot\left\langle \mathbf{F}%
_{on\,q}^{Lorentz}\right\rangle \right]
\end{align}
Here we have used the calculations in Eqs. (45) and (46); \ we also note that
the dot product of $\mathbf{v}_{q}$ \ with the term involving $\mathbf{v}%
_{q}\times\lbrack\nabla_{q}\times\mathbf{A}_{\mu}]$ in Eq. (38) vanishes.
\ The average energy in the magnetic field $\left\langle U_{em\;\mu
q}\right\rangle $ changes because of the changing magnetic moment
$\overrightarrow{\mu}$ and also due to the changing position $\mathbf{r}_{q}%
$\ of the passing charge $q$. \ Just as above in \ Eq. (41), the changing
magnetic moment is associated with an electric field $\mathbf{E}_{\mu
}(\mathbf{r}_{q},t)$\ back at the passing charge which changes the kinetic
energy of the passing charge.
\begin{align}
\frac{dU_{q}}{dt} &  =\mathbf{v}_{q}\cdot\left\langle \mathbf{F}%
_{on\,q}^{Lorentz}\right\rangle =q\mathbf{E}_{\mu}(\mathbf{r}_{q}%
,t)\cdot\mathbf{v}_{q}\nonumber\\
&  =-\mathbf{v}_{q}\cdot\frac{q}{c}\frac{\partial}{\partial t}\mathbf{A}_{\mu
}(\mathbf{r}_{q},t)=-\mathbf{v}_{q}\cdot\left[  \frac{q}{c}\left(  \frac
{d}{dt}\overrightarrow{\mu}\right)  \times\frac{\mathbf{r}_{q}}{r_{q}^{3}%
}\right]
\end{align}
The change in the magnetic field energy associated with the changing position
$\mathbf{r}_{q}$ of the passing charge is compensated by the change in the
kinetic energy (in order $1/c^{2}$) of the orbiting charge of the magnetic
moment. \ This energy change can be written in various forms%
\begin{align}
\frac{d\left\langle U_{\mu}\right\rangle }{dt} &  =\mathbf{v}_{q}%
\cdot\left\langle \mathbf{F}_{on\,\mu}^{Lorentz}\right\rangle =-\frac{q}%
{c}\mathbf{v}_{q}\cdot\overrightarrow{\mu}\times\frac{d}{dt}\left(
\frac{\mathbf{r}_{q}}{r_{q}^{3}}\right)  =\overrightarrow{\mu}\cdot
\frac{\partial}{\partial t}\mathbf{B}_{q}(0,t)\nonumber\\
&  =-\left(  \mathbf{v}_{q}\cdot\nabla_{q}\right)  \left(  \frac{q}%
{c}\mathbf{v}_{q}\cdot\mathbf{A}_{\mu}(\mathbf{r}_{q},t)\right)  \nonumber\\
&  =\left(  \mathbf{v}_{q}\cdot\nabla_{q}\right)  \left\langle \frac
{eq}{2c^{2}r_{eq}}\left(  \mathbf{v\cdot v}_{q}+\frac{(\mathbf{v}%
\cdot\mathbf{r}_{eq})(\mathbf{v}_{q}\cdot\mathbf{r}_{eq})}{r_{eq}^{2}}\right)
\right\rangle \nonumber\\
&  =\left\langle e\mathbf{v\cdot}\frac{q\mathbf{r}_{eq}}{c^{2}r_{eq}^{3}%
}\left(  \frac{-1}{2}\frac{v_{q}^{2}}{c^{2}}+\frac{3}{2}\frac{(\mathbf{v}%
_{q}\cdot\mathbf{r}_{eq})^{2}}{c^{2}r_{eq}^{2}}\right)  \right\rangle
=\left\langle e\mathbf{v}\cdot\lbrack\mathbf{E}_{q}(\mathbf{r},t)-\mathbf{E}%
_{q}^{(0)}(\mathbf{r},t)]\right\rangle
\end{align}
and corresponds to energy delivered to a moving charge by the emf of the
changing magnetic field of the passing charge. \ We notice that it is the
relativistic $v_{q}^{2}/c^{2}$ terms in the electric field $\mathbf{E}_{q}$
which deliver the power to the orbiting charge. \ Thus in the energy
conservation law, the changing magnetic field energy $\left\langle U_{em\;\mu
q}\right\rangle $ is associated with the changing energy of both the magnetic
moment and the passing charge. \ 

\subsection{Center-of-Energy Motion for the Magnetic Moment}

In this section we will discuss the motion of the center of energy of the
magnetic moment from two points of view. \ First we connect its motion to the
motion of the passing charge using the conservation law for the constant
motion of the system center of energy. \ Second we use the particle equations
of motion to obtain what has been called "the equation of motion of the
\ magnet," referring to the center of energy motion of the magnetic moment

The Darwin Lagrangian gives constant velocity to the system center of energy
to order $1/c^{2}$, the same order to which the Darwin Lagrangian is invariant
under Lorentz transformations.\cite{CVV} \ The center of energy
$\overrightarrow{\mathbf{X}}$ to order $1/c^{2}$ involves only rest mass
energy and electrostatic energy%
\begin{align}
\frac{U}{c^{2}}\overrightarrow{\mathbf{X}}  &  =\frac{U_{\mu}}{c^{2}%
}\overrightarrow{\mathbf{X}}_{\mu}+\left[  \frac{U_{em\;eq}}{c^{2}}\left(
\frac{\mathbf{r+r}_{q}}{2}\right)  +\frac{U_{em\;-eq}}{c^{2}}\left(
\frac{\mathbf{r}_{q}}{2}\right)  \right]  +\frac{U_{q}}{c^{2}}\mathbf{r}%
_{q}\nonumber\\
&  =\frac{1}{c^{2}}\left(  U_{\mu}^{[q=0]}-e\mathbf{r}\cdot\mathbf{E}%
_{q}(0)\right)  \left(  \overrightarrow{\mathbf{X}}_{\mu}^{[q=0]}%
+\delta\overrightarrow{\mathbf{X}}_{\mu}\right) \nonumber\\
&  +\left[  \frac{eq}{r_{eq}}\left(  \frac{\mathbf{r+r}_{q}}{2}\right)
+\frac{-eq}{r_{q}}\left(  \frac{\mathbf{r}_{q}}{2}\right)  \right]  +\left[
m_{q}\left(  1+\frac{1}{2}\frac{\mathbf{v}_{q}^{2}}{c^{2}}\right)  \right]
\mathbf{r}_{q}%
\end{align}
where $U_{\mu}^{[q=0]}$ and $\overrightarrow{\mathbf{X}}_{\mu}^{[q=0]}$
correspond to the energy and center of energy of the magnetic moment when the
passing charge is not present. \ When averaged over the orbital motion of the
magnetic moment, the electromagnetic field contribution \ in Eq. (50) yields a
quadrupole contribution, corresponding to the neutrality of the magnetic
moment,
\begin{align}
\left\langle \left[  \frac{eq}{r_{eq}}\left(  \frac{\mathbf{r+r}_{q}}%
{2}\right)  +\frac{-eq}{r_{q}}\left(  \frac{\mathbf{r}_{q}}{2}\right)
\right]  \right\rangle  &  =\frac{-eq}{2}\left\langle \frac{\mathbf{r}}{r_{q}%
}-\frac{(\mathbf{r}\cdot\mathbf{r}_{q})\mathbf{r}_{q}}{r_{q}^{2}}+O\left(
\frac{r^{2}}{r_{q}^{2}}\right)  \right\rangle \nonumber\\
&  =\frac{-eq}{2}\left\langle O\left(  \frac{r^{2}}{r_{q}^{2}}\right)
\right\rangle
\end{align}
since $\left\langle \mathbf{r}\right\rangle =0$, and this contribution
vanishes entirely if we average over $\pm e$ and $\pm\omega$ so as to keep
only the magnetic moment contribution. \ Furthermore, the magnetic momentum
contribution in Eq. (50) can be averaged over the orbital motion to give
(through first order in the interaction perturbation)
\begin{align*}
\left\langle \frac{U_{\mu}}{c^{2}}\overrightarrow{\mathbf{X}}_{\mu
}\right\rangle  &  =\left\langle \frac{1}{c^{2}}\left(  U_{\mu}^{[q=0]}%
-e\mathbf{r}\cdot\mathbf{E}_{q}(0)\right)  \left(  \overrightarrow{\mathbf{X}%
}_{\mu}^{[q=0]}+\delta\overrightarrow{\mathbf{X}}_{\mu}\right)  \right\rangle
\\
&  =\frac{1}{c^{2}}U_{\mu}^{[q=0]}\left(  \overrightarrow{\mathbf{X}}_{\mu
}^{[q=0]}+\delta\overrightarrow{\mathbf{X}}_{\mu}\right)  =\frac{\left\langle
U_{\mu}\right\rangle }{c^{2}}\overrightarrow{\mathbf{X}}_{\mu}%
\end{align*}
since $\left\langle \mathbf{r}\right\rangle =0$. \ It follows that Eq. (50)
becomes%
\begin{equation}
\frac{U}{c^{2}}\overrightarrow{\mathbf{X}}=\frac{\left\langle U_{\mu
}\right\rangle }{c^{2}}\overrightarrow{\mathbf{X}}_{\mu}+\frac{\left\langle
U_{q}\right\rangle }{c^{2}}\mathbf{r}_{q}%
\end{equation}
Now differentiating twice with respect to time and noting that the energies
$U$, $\left\langle U_{\mu}\right\rangle ,$ and $\left\langle U_{q}%
\right\rangle $ are all constant in time through 0-order in $1/c^{2},$ while
$d^{2}\overrightarrow{\mathbf{X}}/dt^{2}=0$, we find%
\begin{equation}
0=\frac{\left\langle U_{\mu}\right\rangle }{c^{2}}\frac{d^{2}\overrightarrow
{\mathbf{X}}_{\mu}}{dt^{2}}+\frac{\left\langle U_{q}\right\rangle }{c^{2}%
}\frac{d^{2}\mathbf{r}_{q}}{dt^{2}}%
\end{equation}
Thus the motions of the centers of energy of the magnetic moment and the
passing charge are coupled together. \ Our equations (52) and (53) here
correspond to Eqs. (14) and (15) in Coleman and Van Vleck's discussion of the
interaction of a point charge and a magnet. \ 

For the magnetic moment alone, the center of energy $\overrightarrow
{\mathbf{X}}_{\mu}$ is defined as
\begin{equation}
\frac{U_{\mu}}{c^{2}}\overrightarrow{\mathbf{X}}_{\mu}=m\left(  1+\frac{1}%
{2}\frac{\mathbf{v}^{2}}{c^{2}}\right)  \mathbf{r}+M\mathbf{R}-\frac{e^{2}%
}{c^{2}r}\left(  \frac{\mathbf{r}}{2}\right)
\end{equation}
where the energy $U_{\mu}$ of the magnetic moment through 0-order in $1/c^{2}$
is
\begin{equation}
U_{\mu}=mc^{2}\left(  1+\frac{1}{2}\frac{\mathbf{v}^{2}}{c^{2}}\right)
+Mc^{2}-\frac{e^{2}}{r}%
\end{equation}
and where we have taken the displacement $\mathbf{R}$ of the large mass $M$ as
small compared to $\mathbf{r}$. \ In the nonrelativistic (0-order $1/c$)
limit, the center of energy $\overrightarrow{\mathbf{X}}_{\mu}^{(0)}$
corresponds to the center of (rest) mass%
\begin{equation}
(m+M)\overrightarrow{\mathbf{X}}_{\mu}^{(0)}=m\mathbf{r}+M\mathbf{R}%
\end{equation}
which in our example has been chosen so $\overrightarrow{\mathbf{X}}_{\mu
}^{(0)}=0.$ \ Furthermore, the center of (rest) mass remains at rest since
differentiating Eq. (56) with respect to time leads to the nonrelativistic
statement regarding the momentum of the magnetic moment%
\begin{equation}
(m+M)\frac{d}{dt}\overrightarrow{\mathbf{X}}_{\mu}^{(0)}=m\mathbf{v}%
+M\mathbf{V}=0
\end{equation}
The 0-order (nonrelativistic) linear momentum of the magnetic moment indeed
vanishes since the internal Coulomb forces within the magnetic moment satisfy
Newton's third law and the nonrelativistic Coulomb forces on the two
oppositely charged particles of the magnetic moment due to the distant point
charge $q$ are equal and opposite in the approximation of Eq. (17).

If we differentiate Eq. (54) for the center of energy of the magnetic moment,
we obtain,%
\begin{equation}
\frac{U_{\mu}}{c^{2}}\frac{d\overrightarrow{\mathbf{X}}_{\mu}}{dt}+\frac
{1}{c^{2}}\frac{dU_{\mu}}{dt}\overrightarrow{\mathbf{X}}_{\mu}=m\left(
1+\frac{1}{2}\frac{\mathbf{v}^{2}}{c^{2}}\right)  \mathbf{v}+M\mathbf{V}%
-\frac{e^{2}}{2c^{2}r}\left(  \mathbf{v}-\frac{(\mathbf{r}\cdot\mathbf{v}%
)\mathbf{r}}{r^{2}}\right)  +\frac{m\mathbf{r}(\mathbf{v}\cdot\mathbf{a}%
)}{c^{2}}%
\end{equation}
The acceleration $\mathbf{a}$ of the orbiting charge is given in Eq. (14) and
the time derivative of the energy is related to the work done by the electric
field of the passing charge $dU_{\mu}/dt=e\mathbf{v}\cdot\mathbf{E}_{q}$.
\ Then averaging over the orbital motion, equation (58) becomes
\begin{align}
\left\langle \frac{U_{\mu}}{c^{2}}\frac{d\overrightarrow{\mathbf{X}}_{\mu}%
}{dt}+\frac{1}{c^{2}}\frac{dU_{\mu}}{dt}\overrightarrow{\mathbf{X}}_{\mu
}\right\rangle  &  =\frac{\left\langle U_{\mu}\right\rangle }{c^{2}}%
\frac{d\overrightarrow{\mathbf{X}}_{\mu}}{dt}+\frac{1}{c^{2}}\left\langle
e\mathbf{v}\cdot\mathbf{E}_{q}^{(0)}\right\rangle \overrightarrow{\mathbf{X}%
}_{\mu}=\frac{\left\langle U_{\mu}\right\rangle }{c^{2}}\frac{d\overrightarrow
{\mathbf{X}}_{\mu}}{dt}\nonumber\\
&  =<m\left(  1+\frac{1}{2}\frac{\mathbf{v}^{2}}{c^{2}}\right)  \mathbf{v}%
+M\mathbf{V}-\frac{e^{2}}{2c^{2}r}\left(  \mathbf{v}-\frac{(\mathbf{r}%
\cdot\mathbf{v})\mathbf{r}}{r^{2}}\right) \nonumber\\
+\frac{\mathbf{r}}{c^{2}}\left[  \mathbf{v}\cdot\left(  -\frac{e^{2}%
\mathbf{r}}{r^{3}}+e\mathbf{E}_{q}^{(0)}(\mathbf{r},t)\right)  \right]   &  >
\end{align}
where we have noted $\left\langle e\mathbf{v}\cdot\mathbf{E}_{q}%
^{(0)}\right\rangle =0$. \ Now combining the terms involving $e^{2}$, and
rewriting the average of $e\mathbf{r}\left[  \mathbf{v}\cdot\mathbf{E}%
_{q}^{(0)}\right]  /c^{2}$ as in Eq. (28), we have%
\begin{align}
\frac{\left\langle U_{\mu}\right\rangle }{c^{2}}\frac{d\overrightarrow
{\mathbf{X}}_{\mu}}{dt}  &  =\left\langle m\left(  1+\frac{1}{2}%
\frac{\mathbf{v}^{2}}{c^{2}}\right)  \mathbf{v}+M\mathbf{V}-\frac{e^{2}%
}{2c^{2}r}\left(  \mathbf{v}+\frac{(\mathbf{r}\cdot\mathbf{v})\mathbf{r}%
}{r^{2}}\right)  \right\rangle -\frac{1}{c}\overrightarrow{\mu}\times
\mathbf{E}_{q}^{(0)}(0,t)\nonumber\\
&  =\left\langle \mathbf{P}_{\mu}\right\rangle -\frac{1}{c}\overrightarrow
{\mu}\times\mathbf{E}_{q}^{(0)}(0,t)
\end{align}
This result (60) corresponds to Eq. (26) of the work by Coleman and Van
Vleck.\cite{CVV} \ Next differentiating Eq. (60) with respect to time so as to
obtain a second derivative of $\overrightarrow{\mathbf{X}}_{\mu}$
\begin{equation}
\frac{\left\langle U_{\mu}\right\rangle }{c^{2}}\frac{d^{2}\overrightarrow
{\mathbf{X}}_{\mu}}{dt^{2}}=\frac{d}{dt}\left\langle \mathbf{P}_{\mu
}\right\rangle -\frac{d}{dt}\left(  \frac{1}{c}\overrightarrow{\mu}%
\times\mathbf{E}_{q}^{(0)}(0,t)\right)
\end{equation}
This equation is sometimes called "the equation of motion for a magnetic
moment."\cite{APV}\cite{Vaidman}

\subsection{The Argument over Hidden Momentum in Magnets}

Because the interaction of a magnet and a passing point charge is so poorly
understood, there can arise certain notions which are used as "explanations"
but are not explored in detail. \ "Hidden momentum in magnets" is such a
notion. \ We will illustrate the situation using our calculations for the
interaction of a point charge and a hydrogen-atom magnetic moment which we
have calculated above.

Because the proponents of the no-velocity-change point of view are so sure
that there is no force back on a charged particle passing a magnet, they also
feel sure that there must be no change in the center of energy of the magnet.
\ Thus\ if the center of energy of the magnet did change position, then
according to our Eq. (53) (and according to Coleman and Van Vleck's Eq. (15)),
the passing charge must accelerate. \ Moreover, there is clearly a possibility
of acceleration for the magnet's center of energy since there is an obvious
magnetic Lorentz force on the magnet given by $\mathbf{F}_{on\,\mu}%
=\nabla(\overrightarrow{\mu}\cdot\mathbf{B}_{q})$. \ Now fundamental classical
theorems connect the force and changes in system momentum so that we must have
$\mathbf{F}_{on\,\mu}=\nabla(\overrightarrow{\mu}\cdot\mathbf{B}%
_{q})=d\left\langle \mathbf{P}_{\mu}\right\rangle /dt.$ \ But our equation
(61) gives an escape from motion for the center of energy of the magnet
because there is a second term in the expression for the acceleration of the
center of energy. \ Thus the proponents of the no-velocity-change point of
view decide that the quantity $-(1/c)\overrightarrow{\mu}\times\mathbf{E}_{q}$
represents a "hidden momentum in magnets" whose change "cancels" the classical
applied force. \ Indeed, a mechanical momentum of the required form is
mentioned in a footnote in Coleman and Van Vleck's work\cite{CVV} and now
appears in an electromagnetism text book.\cite{Griffiths} \ However, no one
who speaks of "hidden momentum in magnets" has ever given any relativistic
calculation which shows how this momentum carries out this cancellation
without continuing changes in the charge and current densities of the magnet.
\ "Hidden momentum in magnets" (as used by the proponents of the
no-velocity-change point of view) seems to be an idea which exists simply to
prevent the motion of the center of energy of a magnet. \ As we see above in
our explicit model of a hydrogen-atom magnetic moment and a point charge,
there is indeed a force back on the passing charge and there is indeed motion
of the center of energy of the magnet. \ Both of these results are contrary to
the claims of the proponents of the no-velocity-change point of view. \ 

\section{Transition to a Multiparticle Magnet}

Experimental observation of the interaction of a magnet and a point charge
(such as in the Aharonov-Bohm phase shift) involves not two-particle magnetic
moments but rather multiparticle magnets. \ We are interested in understanding
the experimental situation based upon the insight gained from the fundamental
interaction involving a two-particle magnetic dipole moment.

Within classical electromagnetism, the transition to a multiparticle system is
most familiar for the electrostatics of polarizable particles. \ In our
calculation above, we found that our magnetic moment oriented in the direction
of the displacement $\mathbf{r}_{q},$ $\overrightarrow{\mu}||\mathbf{r}_{q},$
acted like a polarizable particle producing a back force of magnitude
$F_{on\text{ }q}=q^{2}e^{2}/(m\omega^{2}r_{q}^{6})$ back on the point charge
$q.$ When the polarizability is larger (for example, $m$ is smaller for fixed
$\omega$), then the force back on the distant charge is larger. \ Also, when
we have many polarizable particles present, the force back on the distant
particle does not disappear but rather increases to a well-defined limit.
\ Thus if we consider a dielectric\ wall formed by polarizable particles, then
the mutual interaction among the polarizable particles changes the functional
dependence of the force over toward $F_{on\text{ }q}=q^{2}/(2r_{q})^{2},$
which holds for a conducting wall where the force is independent of the
polarizability in the limit of large polarizability. \ This occurs because
polarizable particles which are next to each other in the wall form electric
dipole moments which tend to cancel the external electric field $\mathbf{E}%
_{q}$ at the position of the other electric dipoles in the wall. \ 

In an analogous fashion, we expect multiparticle interactions within a magnet
to alter the back force on a passing charge found in Eq. (35). \ We note that
the force back at the passing charge $q$ due to our model magnetic moment can
be varied by changing the mass of the orbiting charge while keeping the
magnetic moments fixed. \ When the magnetic moment involves a small mass $m$
(and thus is easily influenced by the external electric field $\mathbf{E}%
_{q})$, the force back at the passing charge is larger, just as is true for a
polarizable particle in the electrostatic situation. \ The most symmetrical
multiparticle arrangement of magnetic moments involves $N$ magnetic moments
arranged around a circle as a toroid with the distant charged particle $q$
located along the axis of the toroid.\ \ The 0-order (nonrelativistic)
electrostatic force on each of the orbiting charges of the toroid due to the
charge $q$ is $e\mathbf{E}_{q}^{(0)}$ just as before, while the back force on
the charge $q$ is now $N$ times as large. Again, in analogy with the
electrostatic situation, we expect that due to multiparticle interactions
within the toroid the back force on a passing charge will not disappear but
rather will increase to a limit. \ Now there will be nonrelativistic
electrostatic forces between the charges of the $N$ magnetic moments. \ Also,
each of the orbiting charges $e$ produces acceleration fields of order
$1/c^{2\text{ }}$which act on all of the other orbiting charges of the
magnetic moment. \ Since the $1/c^{2}$-acceleration fields act on each of the
other orbiting charges of the toroid, the back force on \textit{each} orbiting
charge $e$ increases as the number $N$ of two-particle magnetic moments
increases. \ These acceleration fields always cause forces such as to oppose
any change in the currents of the toroid. \ This corresponds to a
self-inductance effect which increases as $N^{2}$ when there are $N$
current-carrying loops.

It is important to notice that the present situation does \textit{not}
correspond to the elementary mutual-inductance problem of electromagnetism
texts. \ In mutual inductance effects, the self-induced emf is such as to
oppose any change in magnetic flux introduced externally and the magnitude of
the back emf is independent of the current which is flowing in the toroid
winding. \ In our case here, the initial accelerations tending to change the
magnetic flux through the toroid do not arise from any induced emf through the
toroid. Indeed in the limit $\mathbf{v}_{q}=0$ there is no emf at all in the
toroid. \ Furthermore, the back force on the charge $q$ does not behave as in
Lenz's law. \ Rather, the tendency to change the currents of the toroid arises
from Solem's strange polarization associated with the \textit{electrostatic
}field of the external charge $q$ treated as a \textit{uniform} electric field
across each magnetic moment; the change in the magnetic moment is proportional
to the magnetic moment and changes sign with the sign of $q$ as seen in Eq.
(35). \ 

We expect that in the multiparticle limit, the electrostatic interactions
within the toroid will tend to screen the field of the passing charge $q$ out
of the toroid and the back force on the passing charge will be limited by the
magnetic energy of interaction. \ Indeed, calculations for ohmic conductors
suggest that the electric fields of a passing charge are screened out of the
body of the conductor by surface charges while the magnetic fields of the
passing charge penetrate into the body of the conductor.\cite{B1999} \ We note
that if the point charge is held at rest outside a conductor, then the
electric fields of the point charge are screened out of the body of the
conductor by surface charges. \ If the charged particle is moving, we do not
expect this electric-field screening to suddenly disappear. \ On the other
hand, it has been shown that magnetic fields due to moving charges penetrate
into an ohmic conductor giving a time-integral of the magnetic field which is
independent of the conductivity of the materials.\cite{B1999} \ As was
suggested earlier, this is precisely the result which is needed to account for
the Aharonov-Bohm phase shift as a classical lag associated with
energy-related classical forces.\cite{B5}

\subsection{Energy, Momentum, and Forces in the Multiparticle Limit}

Let us now consider the momentum, energy, and forces when a charged particle
$q$ moves with velocity $\mathbf{v}_{q}$ down the axis of a magnet in the form
of a toroid which is initially at rest. \ The screening of the electric field
of the passing charge out of the body of the magnet implies that the electric
field vanishes inside the toroid and therefore there is no significant
contribution to momentum from the electromagnetic field of the form
$\mathbf{P}_{em\,\mu q}$ discussed above, and no significant energy flow
across the magnet. \ It follows from Eq. (36), that now the total system
momentum consists of only two contributions, one each from the magnet and the
passing charge%
\begin{equation}
\mathbf{P}=\mathbf{P}_{\mu}+m_{q}\mathbf{v}_{q}\left(  1+\frac{1}{2}%
\frac{\mathbf{v}_{q}^{2}}{c^{2}}\right)  \text{ \ \ multiparticle limit}%
\end{equation}
The Lorentz forces on the magnet and on the passing charge then satisfy
Newton's third law%
\begin{align}
0  &  =\frac{d\mathbf{P}}{dt}=\frac{d\mathbf{P}_{\mu}}{dt}+\frac{d}{dt}\text{
}\left[  m_{q}\mathbf{v}_{q}\left(  1+\frac{1}{2}\frac{\mathbf{v}_{q}^{2}%
}{c^{2}}\right)  \right] \nonumber\\
&  =\mathbf{F}_{on\,\mu}^{Lorentz}+\mathbf{F}_{on\,q}^{Lorentz}\text{
\ \ \ \ \ \ \ \ \ \ \ \ \ \ \ \ \ \ \ \ \ \ \ \ \ multiparticle limit}%
\end{align}
Furthermore, since the electric field is screened out of the body of the
magnet, the center of energy motion of the magnet in Eq. (61) becomes the
familiar Newton's second law connecting the center of mass motion with the net
Lorentz force
\begin{equation}
\frac{\left\langle U_{\mu}\right\rangle }{c^{2}}\frac{d^{2}\overrightarrow
{\mathbf{X}}_{\mu}}{dt^{2}}=\frac{d}{dt}\left\langle \mathbf{P}_{\mu
}\right\rangle =\mathbf{F}_{on\,\mu}^{Lorentz}%
\end{equation}

The net Lorentz force on the magnet is exactly the original standard classical
magnetic Lorentz force\cite{J5} on the magnet due to the magnetic fields of
the passing charge,
\begin{align}
\mathbf{F}_{on\,\mu}^{Lorentz}  &  =\left[  \nabla_{\mathbf{r}}\left\{
\overrightarrow{\mu}\cdot\mathbf{B}_{q}(\mathbf{r,}t\mathbf{)}\right\}
\right]  _{\mathbf{r}=0}=-\nabla_{q}\left\{  \overrightarrow{\mu}\cdot\left(
q\frac{\mathbf{v}_{q}}{c}\times\frac{(-\mathbf{r}_{q}\mathbf{)}}{r_{q}^{3}%
}\right)  \right\} \nonumber\\
&  =-\nabla_{q}\left\{  \frac{q}{c}\mathbf{v}_{q}\cdot\lbrack\overrightarrow
{\mu}\times\frac{\mathbf{r}_{q}}{r_{q}^{3}}]\right\}  =-\frac{q}{c}\left(
\mathbf{v}_{q}\cdot\nabla_{q}\right)  \mathbf{A}_{\mu}(\mathbf{r}_{q})
\end{align}
where we have written the magnetic field of the charged particle evaluated at
the origin as $\mathbf{B}_{q}=q\mathbf{v}\times(-\mathbf{r}_{q}\mathbf{)}%
c^{-1}r_{q}^{-3}$, have used standard vector identities, have recognized the
magnetic vector potential $\mathbf{A}_{\mu}(\mathbf{r}_{q})=\overrightarrow
{\mu}\times\mathbf{r}_{q}/r_{q}^{3}$ of the magnet at the position of the
charged particle $q$, and have dropped the magnetic Lorentz force
$(q/c)\mathbf{v}_{q}\times\mathbf{B}_{\mu}$ which vanishes for a point charge
$q$ on the axis of a toroid. \ Newton's third law in Eq. (63) for the forces
between the toroid and the passing charge requires that
\begin{equation}
\mathbf{F}_{on\,q}^{Lorentz}=\frac{q}{c}\left(  \mathbf{v}_{q}\cdot\nabla
_{q}\right)  \mathbf{A}_{\mu}(\mathbf{r}_{q})
\end{equation}

While the electric velocity field of a passing charge is screened out of a
good conductor, the magnetic field penetrates into a good conductor with a
time integral which is independent of the conductivity of the ohmic material
of the conductor.\cite{B1999} \ \ Thus the magnetic field energy $U_{em}$
associated with the overlap of the toroid magnetic field and the point charge
magnetic field\cite{B1} is
\begin{equation}
U_{em\,\mu q}=\frac{1}{8\pi}%
{\displaystyle\int}
d^{3}r\,2\mathbf{B}_{q}\cdot\mathbf{B}_{\mu}=q\frac{\mathbf{v}_{q}}{c}%
\cdot\mathbf{A}_{\mu}(\mathbf{r}_{q})
\end{equation}
just what was given for $\left\langle U_{em\;\mu q}\right\rangle $ in Eq.
(45). \ Let us assume that this magnetic field energy is equal to the change
in kinetic energy of the passing charge due to the electric fields from the
changing charge and current densities of the magnet. \ Since the change in
magnetic field energy is of order $1/c^{2}$, we need to consider only the
nonrelativistic approximation to the passing particle kinetic energy. \ Then
we find%
\begin{align}
\frac{1}{2}m_{q}\mathbf{v}_{q}^{2}-\frac{1}{2}m_{q}\mathbf{v}_{q0}^{2}  &
=U_{em\,\mu q}\nonumber\\
m_{q}\mathbf{v}_{q0}\cdot\Delta\mathbf{v}_{q}  &  =\frac{q}{c}\mathbf{v}%
_{q0}\cdot\mathbf{A}_{\mu}(\mathbf{r}_{q})
\end{align}
where $\mathbf{v}_{q0}$ is the velocity of the charged particle $q$ when far
from the magnet where $\mathbf{A}_{\mu}(\mathbf{r}_{q})$ vanishes, and
$\Delta\mathbf{v}_{q}$ is the change in the velocity of the passing charge.
\ Thus we find%
\begin{equation}
m_{q}\Delta\mathbf{v}_{q}=(q/c)\mathbf{A}_{\mu}(\mathbf{r}_{q})
\end{equation}
and the force on the passing charge is therefore%
\begin{equation}
\mathbf{F}_{on\,q}=m_{q}d\mathbf{v}_{q}/dt=m_{q}d(\Delta\mathbf{v}%
_{q})/dt=(q/c)\left(  \mathbf{v}_{q}\cdot\nabla_{q}\right)  \mathbf{A}_{\mu
}(\mathbf{r}_{q})
\end{equation}
exactly as found in Eq. (66) from Newton's third law. \ Thus there is a
certain consistency between our momentum and energy considerations. \ However,
it should be noted that the kinetic energy change for the passing charge is
assumed to be of the same sign as the change in energy of the magnetic field.
\ Energy conservation thus requires that the charges carrying the currents of
the toroid must absorb twice the kinetic energy change of the passing charge.
\ If the currents of the toroid act in a fashion analogous to a battery in
magnetic systems involving mechanical work, then such an energy balance is
consistent with what is found for familiar magnetic systems.\cite{flat} \ We
note that the energy absorbed by the center of mass motion of the magnet is of
order $1/c^{4}$ and hence is negligible, since the recoil velocity of the
center of energy of the toroidal magnet (which was initially at rest) is of
the order of $1/c^{2}$ from Eq. (65).

One should note the difference in perspectives between the analysis given here
in the classical-lag point of view and that suggested by proponents of the
no-velocity-change point of view (those who support the quantum topological
interpretation of the Aharonov-Bohm phase shift). It was pointed out by
Coleman and Van Vleck,\cite{CVV} and repeated above in Eq. (53), that the
accelerations of the centers of energy for the toroid and the passing charge
must be related as in Newton's third law. \ We have assumed that the electric
field of the passing charge is screened out of the magnet, have obtained the
force on the passing charge $q$ by assuming that\ it is the third law partner
of the usual magnetic Lorentz force on the toroidal magnet, and then have
shown that this force is directly related to the energy change in the magnetic
fields which penetrate into the magnet. \ The no-velocity-change point of view
claims that there is no force back on the passing charge, that the magnetic
moment of the magnet does not change, and that the changing electromagnetic
field momentum is associated with "hidden momentum in magnets" whose change
"cancels" the magnetic Lorentz force on the magnet. \ This requires that the
electric field of the passing charge should penetrate into the magnet so as to
give the "hidden momentum," a penetration which seems contrary to the
screening of electric fields by conductors. \ Furthermore, this point of view
tells us nothing about magnetic energy changes between the passing charge and
a toroid. \ 

\section{Discussion}

Although the Aharonov-Bohm phase shift is well known and is now standard in
all the recent quantum mechanics texts, most physicists seem unaware of the
long-standing controversy regarding the interpretation of the phase shift.
\ In 1959, Aharonov and Bohm\cite{AB} solved the Schroedinger equation and
predicted their phase shift. \ The phase shift has been observed
experimentally.\cite{Chamb} \ Aharonov and Bohm attracted attention to their
phase shift by claiming that their predicted phase shift occurred in the
absence of classical electromagnetic forces and velocity changes and
represented a new quantum topological effect with no analogue in classical
theory. \ There is no experimental evidence for this claim. \ Indeed, the
interpretation has aroused controversy. \ Most of the initial controversy
regarding the Aharonov-Bohm phase shift centered on a distraction, whether or
not the shift was caused by stray magnetic fields outside the solenoid or
toroid. \ This aspect of the controversy has been removed by the toroidal
experiments of Tonomura\cite{T}\ \ which allow very little stray magnetic flux.

The suggestion that the Aharonov-Bohm phase shift might be based upon a
classical lag effect involving classical electromagnetic forces and velocity
changes (the suggestion repeated here) depends upon our understanding of
classical electromagnetism. \ The conventional attitude regarding the
Aharonov-Bohm phase shift is best stated by Aharonov, Pearle, and
Vaidman:\cite{APV} "In the Aharonov-Bohm effect it is obvious that the
electron is not subject to any electromagnetic force, because the magnetic
field lies wholly within the filament and so is zero at the electron's
location." \ This naive statement omits the crucial possibility of induced
charge or current densities in the magnet leading to forces back on the
passing charge. \ Indeed, induced currents do lead to forces back on passing
charges; the phase shifts may well arise from classical lag effects. \ 

In the 1970s, it was suggested that the possible influence of the
electromagnetic fields of the passing charge could be removed by surrounding
the solenoid or toroid by a conductor which would screen out the
electromagnetic fields.\cite{pre} Experiment showed that the phase shift
persisted even when the solenoid was surrounded by a conductor.\cite{T}
\ However, it was realized that although electric fields are indeed well
screened by a conductor, magnetic velocity fields penetrate into an ohmic
conductor (and also into superconductors at high frequencies) in a form which
is completely different from the skin-depth behavior of wave fields, and
indeed there is an invariant time integral which has precisely the correct
form to account for the Aharonov-Bohm phase shift as an energy-related lag
effect based on classical forces.\cite{B1999} \ The experiments to date do no
not remove the possibility of a classical electromagnetic basis for the
Aharonov-Bohm phase shift.\cite{B5} \ In addition, it was pointed out that
electrostatic forces can give interference pattern shifts which take exactly
the same form as the Aharonov-Bohm phase shift.\cite{B2} \ Matteucci and Pozzi
confirmed this experimentally in 1985.\cite{MP}\ 

In 1984, Aharonov and Casher\cite{AC} suggested a second phase shift, this
time for a magnetic moment passing a line charge, which they claimed was the
duel of the Aharonov-Bohm phase shift and again occurred in the absence of
classical forces and velocity changes. \ \ However, it was pointed out that
conventional classical electromagnetic theory clearly predicted a force on a
passing magnetic moment treated as a current loop, and Newton's second law
suggested a lag effect.\cite{B4} \ To counter this observation, Aharonov,
Pearle and Vaidman\cite{APV} introduced a new analysis for the interaction of
a magnetic moment and a point charge, and claimed that the magnetic moment,
although indeed experiencing a net Lorentz force, nevertheless moved as though
it experienced no forces whatsoever, because of changes in "hidden momentum in
magnets" cancelling the applied Lorentz force. \ 

For the Aharonov-Bohm phase shift, the Aharonov-Casher phase shift, and the
Shockley-James paradox, the heart of the controversy and paradox involves the
interaction between a point charge and a magnetic moment through order
$1/c^{2}.$ \ Although the literature of the Aharonov-Bohm phase shift is full
of statements about the interaction which claim to exclude any possibility of
an explanation based upon classical electromagnetic forces\cite{P}, the claims
often depend upon nonrelativistic models\cite{PTT} or point to familiar
effects, such as aspects of mutual inductance, which indeed will not give the
desired behavior,\cite{mut}\ but overlook the 0-order forces on the charges of
the magnet because the magnet is neutral. \ Coleman and Van Vleck have treated
the interaction consistently relativistically using the Darwin Lagrangian.
\ In the present work, we have followed the Darwin Lagrangian analysis. \ We
have modeled the magnetic moment as a classical hydrogen atom interacting with
the passing charge through the Darwin Lagrangian, and have noted particularly
the nonrelativistic behavior of the magnetic moment pointed out by Solem.
\ The model is unambiguous in its prediction of classical electromagnetic
forces, energies, and changes of the center of energy. \ \ It is the 0-order
accelerations which cause electric fields in order $1/c^{2}$ which act
strongly on the passing charge. \ 

The transition to a multiparticle limit still allows ambiguities. \ However,
the assumption that in this limit the electric fields are screened out of the
magnet while the magnetic fields penetrate into the magnet both fits with what
is known for ohmic conductors and also allows for a consistent treatment of
the conservation laws of relativistic theory. \ The discussion given here
represents a refutation of the suggestions of Aharonov, Pearle, and Vaidman
regarding the role of "hidden momentum in magnets" and confirms the
semiclassical calculations of both the Aharonov-Bohm and Aharonov-Casher phase
shifts based upon classical lag effects.\cite{B3}\cite{B4} \ What is needed
now are experiments to test whether or not the Aharonov-Bohm and
Aharonov-Casher phase shifts occur in the presence or absence of velocity
changes for the passing particles.\cite{BCC}

\subsection{Acknowledgement}

I wish to thank Professor Joel Gersten for a number of helpful discussions.


\begin{thebibliography}{99}
\bibitem {nature}The absence of any feedback signal seems surprising. \ It is
not the way nature usually operates.
\bibitem {Momentum}The constancy of the system momentum $\mathbf{P}$ with
respect to time can be checked by differentiating Eq. (36) with respect to
time and then using the equations of motion (11)-(13).
\bibitem {energy}The constancy of the system energy $U$ with respect to time
can be checked by differentiating Eq. (44) with respect to time and then using
the equations of motion (11)-(13).
\bibitem {mut}The changing magnetic flux through the magnetic moment (or
toroid) due to the point charge's magnetic field causes a back force which is
of the order of $1/c^{4}$, (since it does not begin with accelerations due to
the Coulomb force), is independent of the magnetic moment of the toroid, and
(following Lenz's law) is always such as to try to slow the charge. The force
discussed in Eq. (35) here is totally different, always acting to both
accelerate and decelerate the point charge with the time order of the
acceleration and deceleration depending upon the orientation of the magnetic
moment and the sign of the charge.
\end{thebibliography}
\end{document}